\documentclass[sigplan,authorversion=true]{acmart}

\usepackage{booktabs} 

\usepackage{amsmath}
\usepackage{amsthm}

\usepackage{algorithm,algpseudocode}

\DeclareMathOperator*{\argmin}{\arg\!\min}
\newcommand{\tx}{\mathrm{tx}}

\usepackage[font=footnotesize,labelfont=bf]{caption}

\copyrightyear{2017}
\acmYear{2017}
\setcopyright{acmlicensed}
\acmConference[SERIAL'17]{SERIAL'17: ScalablE and Resilient
InfrAstructures for distributed Ledgers}{December 11--15, 2017}{Las
Vegas, NV, USA}
\acmPrice{15.00}
\acmDOI{10.1145/3152824.3152826}
\acmISBN{978-1-4503-5173-7/17/12}

\begin{document}
\title[Economic Analysis of Routing in Payment Channel Networks]{Towards an Economic Analysis of Routing in Payment Channel Networks}

\author{Felix Engelmann}
\author{Henning Kopp}
\author{Frank Kargl}
\affiliation{%
  \institution{Institute of Distributed Systems, Ulm University}
  \streetaddress{Albert-Einstein-Allee 11}
  \city{Ulm}
  \state{BW}
  \postcode{89081}
  \country{Germany}
}
\email{first name.surname@uni-ulm.de}

\author{Florian Glaser}
\author{Chri\-st\-of Weinhardt}
\affiliation{%
  \institution{Information and Market Engineering,\\ Karlsruhe Institute of Technology (KIT)}
  \streetaddress{Fritz-Erler-Straße 23}
  \city{Karlsruhe}
  \state{BW}
  \postcode{76131}
  \country{Germany}
}
\email{first name.surname@kit.edu}

\renewcommand{\shortauthors}{F. Engelmann, F. Glaser, H. Kopp et al.}

\begin{abstract}
Payment channel networks are supposed to
overcome technical scalability limitations of blockchain
infrastructure by employing a special overlay
network with fast payment confirmation and only sporadic settlement of netted
transactions on the blockchain.
However, they introduce economic routing constraints that limit
decentralized scalability and are currently not well understood.
In this paper, we model the economic incentives for participants in payment channel networks.
We provide the first formal model of payment channel economics and analyze how the cheapest path can be found.
Additionally, our simulation assesses the long-term evolution of a payment channel network.
We find that even for small routing fees, sometimes it is cheaper to
settle the transaction directly on the blockchain.

\end{abstract}




\maketitle

\section{Introduction}
Current public blockchain systems such as Bitcoin or Ethereum maintain their
secure, unique global state property by high levels of data replication.
Thus they do not scale with respect to transaction throughput, since
the coordination and communication overhead for replication
significantly reduces performance.
The main bottleneck is the possible number of transactions per second,
that is, changes that can be written to the underlying blockchain database.
One of the most promising approaches to make blockchains scalable are
off-chain state channels~\cite{mccorry2016channelsurvey}.
The approach builds upon the idea of setting up bilateral connections,
so-called channels, between pairs of nodes of a blockchain network.
The blockchain is only used to open and close the channels and
resolve disputes about the current netted amount of previous bilateral transactions.
Off-chain state channels are useful, for example, in the context of
bilateral payments that can take place within such a channel.
Signed transactions can be sent between the parties in both directions and
a bilateral ledger is updated with every transaction.
Only the final state, i.e., the result of netting all transactions of that
bilateral ledger, is persisted in the blockchain if one of the participants
decides to close the channel.
These state channels concerning payments are referred to as payment channels.
Payment channels lead to a higher transaction throughput, since only netted transactions
are written to the blockchain instead of intermediate states.
Payments between nodes that have no direct bilateral connection are
routed through the network of bilateral channels of other nodes.

Payment channels were, up to now, considered mostly from a technical
perspective in terms of saving bandwidth, preventing blockchain
bloat, and enabling cheating resistance.
However, the influence of the economic perspective on usability and
feasibility of payment channels was at most considered as an afterthought.
This is similar to the case of IPv6 where development focused on engineering
aspects instead of how to incentivize protocol adoption,
leading to the situation that IPv6 is not widely used in practice~\cite{rfc7305}.

Although several payment channel protocols have been proposed,
none of them is tested or fully implemented~\citep{mccorry2016channelsurvey}.
However, previous studies already pointed out the technical challenges
of routing in payment channels~\citep{prihodko2016flare}.
Finding a path through the dynamic, unstructured network topology of a
payment channel network is a technically demanding problem.
This is especially the case if the network grows to sizes that are
reasonable for day-to-day payments.

All current payment channel protocol proposals require an initial
deposit by both parties who set up a bilateral channel.
This technical requirement results in capital binding at the opening of a payment channel
for the duration of the lifetime of the channel.
Capital binding economically implies the necessity for remuneration in form of fees.
Such an economic perspective has been the focus of the work of
Miller~et~al.~\cite{miller2017sprites},
who propose changes to a payment channel protocol based on Ethereum in order to
lower the maximum duration of the settlement.
However, the open question regarding technical and economical feasibility remains unanswered.

Against this background, we analyze a generic payment channel protocol regarding its technical and economical feasibility.
The overall goal is to answer the following research question:
What are the implications of economic incentives for routing
in and usage of payment channel networks?

The remainder of the paper is structured as follows.
In the next section we explain the necessary background of payment
channels. \Cref{s_challenges} discusses the main challenges of routing
in payment channel networks. We discuss how to compute the cheapest
route in~\Cref{s_ourapproach}. In~\Cref{s_evaluation} we evaluate the
routing mechanics in payment channels relying on a simulation. We conclude
with~\Cref{s_conclusion}.

\section{Background}\label{s_background}
For Bitcoin two main proposals of bidirectional payment
channels currently exist: duplex micropayment channels (DMC)~\citep{decker2015duplex}
and the Lightning network~\citep{poon2015lightning}.
A good overview of the technical details of payment channels is given
by McCorry et~al.~\citep{mccorry2016channelsurvey}.
Both mechanisms consist of three phases:

\textbf{Setup Phase:}
        In the first phase, two parties set up a bilateral
        payment channel between them. Either party needs to allocate an amount
        of coins to prove that it is able to pay the subsequent
        transactions. This amount is called the capacity of the
        channel. It can differ in the two directions and is the
        maximum total amount that can be routed in each direction.

\textbf{Trading Phase:}
        In the second phase, each of the two parties can send multiple signed
        transactions to the other party, where subsequent transactions
        have priority over the previous ones.
        This enables to retain the strong security properties of the
        blockchain since no party can cheat by committing an older transaction
        to the ledger.
        In DMC, priority of transactions is achieved by a timelocking
        mechanism which enables a later transaction to be spent earlier than
        previous transactions.
        In Lightning, the priority of transactions is implemented by
        continuously revoking the old transactions. If one party includes a
        revoked transaction in the blockchain, the other party can issue a
        penalty to claim all bitcoins in the channel.

 \textbf{Settlement Phase:}
        The third phase of current payment channel proposals consists
        of closing the channel. This can be necessary if the channel is not
        needed anymore or if the initial capacity of the channel is depleted,
        i.e., one participant has received all the money the other party has
        initially deposited and thus no further transactions are possible in
        this direction. Depending on the implementation, capacity can be replenished.

The two already mentioned proposals, DMC and Lightning, are running on an
unspent transaction output (UTXO) based blockchain, that is the Bitcoin
blockchain.
In contrast, a third implementation currently under development---the Raiden Network (http://raiden.network)---relies
on the account state-based Ethereum blockchain.
The three general phases described above also occur in Raiden, however,
the technical implementation differs significantly.
In UTXO blockchains, smart contracts are special output conditions
specified in program code which is attached to transactions.
In Raiden, a channel is a smart contract that contains the channel
setup properties and is represented by a separate account
which is created when the smart contract code is deployed on the Ethereum blockchain.
Nonetheless, the functionality provided by Raiden is comparable to DMC
and Lightning, at least if compared on the abstract level of investigation in this work.

Raiden additionally implements auxiliary smart contracts which are
deployed to the public Ethereum blockchain to support, for example,
the tracking of existing channels and hence facilitate the routing of a payment
over multiple hops.
Raiden provides additional features which facilitate recovering the
state of off-chain channels when restarting a node.
It is worth noting here, that nodes involved in a transaction must be
online during the time the transaction is performed.
The closing of channels could, however, be supported by third parties who
monitor the closing of channels on behalf of the party that
is offline when the opposite party closes the channel.

All three---Raiden, DMC, and Lightning---support routing of transactions over multiple
hops.
A user $A$ can send a payment via a node $B$ to $C$ if payment
channels are set up between $A$ and $B$, and between $B$ and $C$.
To achieve this, $A$ sends its payment to $B$ and then $B$ sends the corresponding
amount to $C$, possibly deducting a routing fee.
These two transactions use hashed timelocks (HTLC) to enforce
dependency~\citep{decker2015duplex}.
Thus, either none or both of the payments are processed.

Since each node needs to have sufficient capacity for its payments,
routing a payment of size $X$ along each of the $n-1$ edges between
$n$ nodes requires each channel to have a capacity of at least $X$
units of the underlying (crypto)currency.
Consequently, in total $\mathcal{O}(Xn)$ capital needs to be locked to
route a transaction of size $X$ over $n$ hops.
In addition to the capacity constraints, economic routing
considerations arise, since
intermediate nodes may demand a premium for their cost of
capital binding and operating a node in the network.
Thus, senders aim to minimize their cost of transferring their
transactions instead of the number of hops.

To our best knowledge, optimal routing algorithms in payment channel
networks are currently a neglected problem and demand further research.
While the work of Prihodko et al.~\citep{prihodko2016flare}
provides a first step in this direction by proposing a routing algorithm
for establishing routes between transaction partners, it fails to take
the full financial dimension into account.

\section{Challenges}\label{s_challenges}
Beyond problems of traditional routing scenarios, like changing
topologies, payment channel networks need to deal with additional
economically induced challenges.

First, the state of payment channels is changing with every transfer due to updated balances.
Hence, the capacities of the network graph's edges are highly
dynamic as a result of the design of payment channels.
This affects the transaction amount which can be routed over a
channel.
In addition, every node along the route of a transaction is
remunerated by taking a fee for every forwarded transfer.
These fees can change over time at the discretion of every node.
Hence, not only the capacities are changing over time, but also the costs per node.
Increasing fees at a node may lead to paths becoming
invalid, since the capacity is not sufficient anymore.
Further, after a first transaction has been routed, the same path may
become invalid for a second, identical transaction due to channel
updates.

A common assumption regarding this problem is that there will be roughly equal
amounts of payment in both directions at every node.
We leave possible real world scenarios in which this assumption holds to the imagination of the reader.

Second, in all three proposed systems, DMC, Lightning,
and Raiden, nodes are allowed to open channels arbitrarily.
This results in an unstructured network where
the identifier of a node is not related to its location in the network.
That is, there is no data structure or node-ID assignment mechanism
(like finger tables based on data hashes in P2P networks) to guide the
transaction hierarchically towards its recipient.
On the other hand, however, a full view of the graph of active channels can be obtained either by
gathering all opening and closing transactions (in UTXO systems) or by inspecting
the account of the smart contract that is tracking open channels (account-based systems).
Hence, we can assume that a full view of the graph is available.

Third, channels need to be equipped with enough funding at setup time in order
to route a transaction of certain size through a channel.
If too many transaction are routed through a channel in the same
direction, the channel capacity will be depleted and no more
transactions can be routed in this direction.
One possible solution to the problem of channel depletion could be
similar to multipath TCP where transactions, which cannot be routed along
a single path, are split into multiple smaller transactions which
take different routes to their recipient.
Another possible solution could be the adjustment of routing fees in
such a way that the fees cover the cost of the capacity imbalance, and
thus lead to a balancing of the channel by market mechanisms.
However, current implementations do not explicitly incentivize or
provide decision support for setting routing fees appropriately.

If routing fails, a regular blockchain transaction
is still a feasible option.

In combination, these properties of current payment channel proposals
make routing in payment channels more complicated than routing in other
P2P networks.

\section{Our Approach}\label{s_ourapproach}
In this section we introduce our formal notation, and suggest an improved solution to
the computation of the cheapest path assuming that transactions are
not split but routed as a whole.

\subsection{Notation}
A payment channel network is composed of nodes $v_i \in V$  that are linked
to each other via bi-directional payment channels.
We write $G = (V, E)$ for the directed graph spanned by the channels between nodes.

A channel, i.e., an edge, between node $v_i$ and node $v_j$  is denoted by $e_{ij}$
with $e_{ij} \in E$. Since the graph is directed, the index is
ordered, i.e., $e_{ij}\neq e_{ji}$ for all $i$, $j$.
The capacity in the channel $e_{ij}$ between node $v_i$ and $v_j$ is denoted by $\omega_{ij}$.
This is the aggregated volume of payments that can be made along $e_{ij}$.
Note that $\omega_{ij}$ may be different from the capacity in the other direction $\omega_{ji}$.
Over time the capacities in the channels change according to the payments made
in the channel as described previously.

Each node $v_i$ which routes a transaction $\tx$ to $v_j$ demands a
financial reward $\rho(e_{ij}, \tx)$ depending on the edge $e_{ij}$ and
attributes of the transaction $\tx$.
In practice it will most likely depend on the size of the transaction in bytes,
since this corresponds to used bandwidth, or it will depend on the transaction
amount $\alpha(tx)$, since the higher the transaction amount, the bigger the imbalance
which is created in the channel.

Routing a transaction $\tx$ along $e_{ij}$ decreases $\omega_{ij}$
by the transaction amount $\alpha(\tx)$ and increases the inverse capacity
$\omega_{ji}$ by the same transaction amount. However, the sum
$\omega_{ij}+\omega_{ji}$ is always constant.

In order to minimize routing costs for a single transaction, agents need to send a
transaction $\tx$ via the cheapest path through the network.
We define the cheapest path $path(\tx)$ of the transaction $\tx$ with
sender $s$ and receiver $r$ as the directed path $\mathcal{P}$ from
$s$ to $r$, where
\(\sum_{e_{ij}\in\mathcal{P},\,v_i\neq s}\rho(e_{ij}, \tx)\)
is minimized.
Note that we exclude the fees of the sender for routing its own
transaction in the sum.

\subsection{Cheapest Path as a Linear Program}
In the following we describe a linear program whose result yields the
cheapest path $path(\tx)$ for a single transaction $\tx$.

We introduce the variables $x_{ij}$ with the following meaning
\[
    x_{ij}:=
    \left\{
    \begin{array}{ll}
        1 & \text{, if $\tx$ is routed along $e_{ij}$}\\
        0 & \text{, else}
    \end{array}\right.
\]

Regarding the constraints we have \emph{preservation constraints} as well as
\emph{capacity constraints}.
The \emph{preservation constraints} assure that there is a path between
the sender $s$ and the receiver $r$. For each node with an incoming
path edge, there is also an outgoing path edge, except if the node is the
sender or the receiver.
That is, for all nodes $v_i$ we require that
\[
    \sum_j x_{ij}-\sum_j x_{ji} =
    \left\{
    \begin{array}{ll}
        1 & \text{, if $v_i$ is the sender $s$}\\
        -1& \text{, if $v_i$ is the receiver $r$}\\
        0 & \text{, else}
    \end{array}\right.
\]

The \emph{capacity constraints} enforce that the capacities along the path
suffice to route the transaction. For all subsets $S$ of the nodes which
include the sender, the sum of the capacities of the outgoing channels
need to support the routing fee outside the set $S$, and the
transaction amount for each outgoing edge.
Formally, for all $S\subset V$ with $s\in S$ we require that
\[
    \sum_{i:\,v_i\in S} \sum_{j:\,v_j\not\in S} \omega_{ij} x_{ij} \geq
    \alpha(\tx) + \sum_{i:\,v_i\not\in S}\sum_{j:\,v_j\not\in S}\rho(e_{ij}, \tx) x_{ij},
\]
where again the symbol $\alpha(\tx)$ denotes the amount of the transaction.
If there is exactly one outgoing edge and no incoming edge to the subset $S$,
the bound is tight.

In summary, we receive the following linear program to compute a cheapest route.

\[ \min \sum_i\sum_j \rho(e_{ij}, \tx) \cdot x_{ij} \]
s.t.
\begin{equation*}
\setlength\arraycolsep{1.7pt}
    \begin{array}{rcl}
        \sum\limits_j x_{ij}-\sum\limits_j x_{ji} &=&
            \left\{
            \begin{array}{ll}
                1  & \text{, if $v_i$ is the sender}\\
                -1 & \text{, if $v_i$ is the receiver}\\
                0  & \text{, else}
            \end{array}\right.\quad\text{for all $i$}\\
            \sum\limits_{i:\,v_i\in S} \sum\limits_{j:\,v_j\not\in S} \omega_{ij} x_{ij} &\geq& \alpha(\tx) + \sum\limits_{i:\,v_i\not\in S}\sum\limits_{j:\,v_j\not\in S}\rho(e_{ij}, \tx) x_{ij} \\
            & & \text{ for all $S\subset V$ with $s\in S$}
    \end{array}
\end{equation*}

Since solving the linear program is computationally difficult, due to
the number of constraints, we decided to take another approach for
the practical computation of the cheapest path as is outlined next.

\subsection{Practical Computation of the Cheapest Path}

In this section we describe our routing algorithm for finding the
cost minimal route in the network for a given transaction as pictured
in~\Cref{a_moddijkstra}.
In order to tackle this problem we make the following assumptions.

We assume that there is a global view of the network of currently open channels.
This is in line with our argument that the current channel connectivity can be
obtained from the public state of the blockchain.
However, only the connection and the initial capacity are publicly visible.
Neither capacities nor channel fees are available in the blockchain. 
Our approach is therefore intended
for research on economic properties of payment channel networks.

We further assume that there is at most one payment channel between each pair of
nodes. While this may not always be the case, we are able to abstract
multiple channels into a single channel of higher capacity.

Lastly, we assume that transactions are not split into
multiple smaller transactions, that is always the full amount is routed along one path.

A common approach to solve shortest-path problems is the well-known Dijkstra algorithm.
However, the algorithm does not work in our use-case,
since it only yields an optimal solution if the visited part of
the network does not change during the calculation.
Since we include routing fees, by extending the path we increase the amount that needs to be
routed at each node on the path, thereby destroying the matroid property.
Hence already visited edges may have insufficient capacity for the transaction and
its routing fees depending on the remaining route and
thus are unable to route the transaction.
Consequently, a path with sufficient capacity cannot be calculated by removing
all edges with a weight smaller than the transaction amount and applying the Dijkstra algorithm.

To keep the matroid property and be able to compute a cheapest route efficiently,
we reverse the graph so that edges change direction but keep their capacity.
We try to find a path by following the route from the receiver to the sender
and add the routing fees to each path.
Already visited nodes and edges will not be removed because of
capacity violations as they all have an admissible path towards the receiver.
When starting with the required amount at the receiver,
the routing fees add up towards the sender,
who learns how much more has to be transmitted to cover the routing fees.

Overall, this approach enables us to find the cheapest path while facing
all of the outlined challenges. 
Consequently, we can obtain insights about the economic applicability of payment channels
in general, without having to worry about the routing complexities in detail.

\begin{algorithm}
  \footnotesize
  \caption{Our algorithm for finding a cheapest spanning tree from the receiver}
  \label{a_moddijkstra}
  \begin{algorithmic}[1]
  \Require the graph $(V,E)$, recipient $r$, transaction $\tx$.
  \Ensure a cheapest spanning tree $\mathcal{T}$ for transactions to $r$.
  \State $Q\gets V$
  \State $\mathcal{T} \gets \emptyset$
  \State $cost(r)\gets 0$,\quad $cost(v)\gets \infty$ for all $v\neq r$
  \While{$Q\neq\emptyset$}
      \State $v_i\gets\argmin\{cost(v),\,v\in Q\}$
      \State $Q\gets Q\setminus\{v_i\}$
      \ForAll{$e_{ji}\in E$}
          \If{$cost(v_j)+\rho(e_{ji},\tx)+\alpha(\tx)\leq \omega_{ji}$}
              \If{$cost(v_i)+\rho(e_{ji},\tx) < cost(v_j)$}
                  \State $cost(v_j)\gets cost(v_i)+\rho(e_{ji},\tx)$
                  \State $path(v_j)\gets e_{ji}$
              \EndIf
          \EndIf
      \EndFor
  \EndWhile
  \State $\mathcal{T} \gets \emptyset$
  \ForAll{$v\in V$}
      $\mathcal{T}\gets T\cup\{path(v)\}$
  \EndFor \\
  \Return $\mathcal{T}$
\end{algorithmic}
\end{algorithm}

\section{Evaluation}\label{s_evaluation}
We implemented a simulator to obtain insights on the performance and
behavior of a payment channel network at larger scale.
This has various benefits over using a real payment channel implementation,
such as elimination of cryptographic overhead and ease of statistical analysis.

The capabilities of our simulator comprise nodes which can open and
close channels with a specified initial volume to another node,
as well as the accounting of the channels.

As the characteristics of the routing fee have a strong influence on the network,
we allow for each node to define its own fee structure in our network simulator.
Initially we use a 0.5\,\% fee of the routed transaction amount,
but adapt the fee reciprocal to the imbalance in the channel.

As our goal is to evaluate the performance and applicability of payment channels
for a replacement of current means of payment, we based our measurements on real world statistics.
To generate our transaction volumes, we fitted a lognormal
distribution to the transaction statistics that were compiled in 2012 and published by the US
Federal Reserve~\cite{fed2013fedstudy}.
According to the data we set a mean of 2.95 and a standard deviation
of 1.2 for a single payment's volume.
These volumes are then transmitted between two nodes chosen uniformly at random.
This assumption is in favor of the implicit assumption of current proposed systems that enough payments
are routed between individuals such that the transfers cancel out---at least on average.
The assumption is implicit as there are no mechanisms incorporated that address channel exhaustion.
Translated into a real world scenario, this would imply that banks are not existing and
transactions take place directly between service providers and consumers in everyday life.
Otherwise, channels would have to be replenished frequently
either due to routed transfers or due to own transactions.

The topology of channels is created by a random graph model known to yield
node degree distributions that resemble real world social networks.
We choose the Barab\'{a}si-Albert graph generation
algorithm~\cite{albert2002statistical} to account for the
power law distribution of node degrees which is common in social networks.
An often mentioned application scenario is ``banking the underbanked'' which
refers to rural areas in development countries. In this scenario, for example,
these social structures can be assumed to be realistic even in the context of retail payments.

\textbf{Results:}
Apart from the initial channel network and the transactions' participants and volumes,
the simulation has multiple other degrees of freedom.
The initial balance of the nodes, if too low, can influence the simulation
as it might not provide enough funding to open a channel or handle a transaction directly on the blockchain.
We avoid these effects by always setting a high enough initial balance.
The capacities of the channels themselves can lead to constraint failures.
At the beginning of our simulation, all channels are funded with an equal amount allocated in both directions.
By design, a transaction larger than the sum of both balances can never be routed along an edge.
Therefore, all larger transactions need to be settled on the blockchain.
For interactions with the blockchain, i.e., channel openings, closings, and direct transactions, we charge a constant fee of 0.41.

\begin{figure}
  \includegraphics{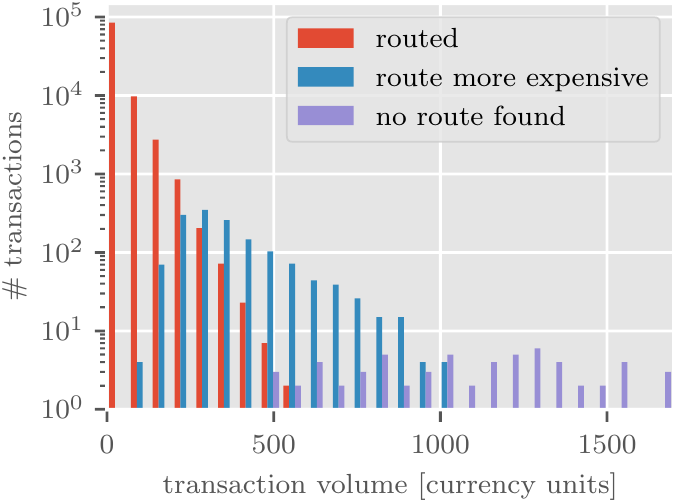}
 \caption{Transactions histogram and methods of settlement for 100\,000 transactions, 1000000 initial node balance, 1000 nodes connected according to a Barab\'{a}si-Albert graph with parameter $m=2$ and 1000 funding from each side in the channels. The routing fees are 0.5\% with an imbalance factor.
A total of 99922 transaction have been routed, among them 1453 where a route exists but is more expensive than a direct transaction and 78 transactions that were settled directly on the blockchain.
 }
 \label{fig:txhist}
\end{figure}
\begin{figure}
   \includegraphics{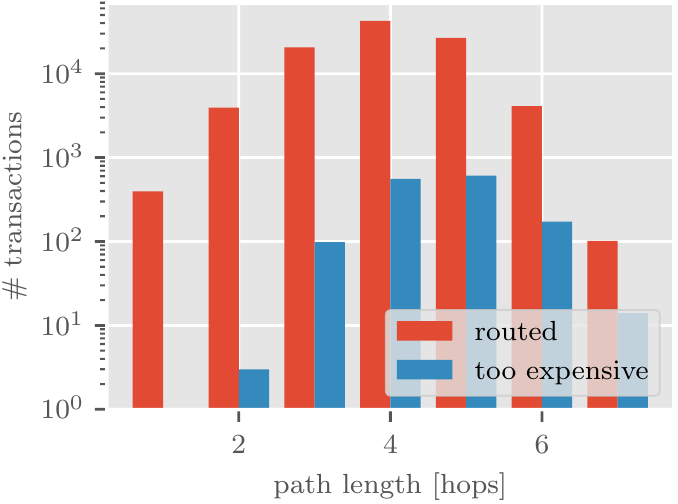}
  \caption{Transactions' path lengths histogram and method from the simulation of Fig. \ref{fig:txhist}}
  \label{fig:lens}
\end{figure}

\begin{figure}[ht!]
  \includegraphics{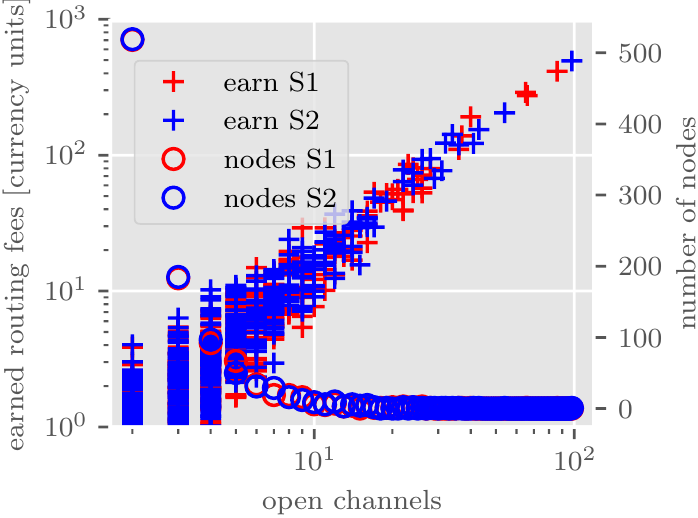}
 \caption{Routing fee earnings histogram for the same parameters as in \Cref{fig:txhist} and two simulation runs.
 The left axis shows the amount a node with the indicated number of open channels earned and to demonstrate how many nodes of this degree appeared in the graph, the right axis shows their distribution.}
 \label{fig:earnings}
\end{figure}

\textbf{Discussion:}
In the transaction histogram \Cref{fig:txhist} the number of blockchain transactions is nearly constant from 500 up to 1000 currency units.
These are transactions from nodes with channels where the channel capacity is not sufficient for the transaction.
Above 1000 currency units, blockchain transactions dominate as a
nearly balanced channel with a funding of 1000 is unable to route these amounts.
The majority of transactions with an amount smaller than 27 units are routed
through channels because with an average route length of 4 hops (3 fees)
the total routing fees are on average lower than the fee for a transaction on the blockchain.
Transactions with more than 82 units are only profitable using direct channels without routing,
except when they help to balance a channel and therefore have a reduced routing fee.
This can be seen in the peak in transaction routes which are possible but more expensive than a blockchain transaction.

Regarding the path length distribution in \Cref{fig:lens}, one hop transactions are always
cheaper than blockchain transactions, as the sender itself charges no fee.
The too expensive routes of low length stem from high volume transactions where the
proportional fee surpasses the constant fee of a blockchain transaction.
For paths with more than 7 hops, the transaction amount has to be lower than 9 units
to be profitably routed, so a larger share of valid routes is more expensive than the constant fee.

As shown in \Cref{fig:earnings}, the earned routing fees of a node increase roughly exponential
with the number of channels and therefore with locked capital.
This means that the revenue is exponential to the invested capital.
To handle all 100\,000 transactions, the 1000 nodes together spent 5576 units
to transfer a total of over 3.9 million units whereas without the payment channels, they would have spent 41\,000 units.
The cost of transactions reduces only slightly for well connected nodes as they have more direct neighbours
to exchange high volume transactions with.


\section{Conclusion \& Future Work}\label{s_conclusion}
Routing in payment channels is a central problem for solving the
scalability issue of blockchains by a state channel approach.
We show that routing in payment channels is more complex due to
economic-technical (econ-technical) constraints.
First, capacity constraints limit
the size of transactions that can be routed and induce capital binding.
Second, channel depletion intensifies capacity constraints and is not addressed by current proposals.
Third, current constraints increase costs of usage,
especially if common blockchain transactions are used as fallback option.
A naive solution would be centralization in form of routing hubs.
However, this would introduce a system that is orthogonal to the intention
and sole purpose of using blockchain systems: decentralization.

We are working on an evolutionary topology construction algorithm
to mitigate economic issues as well as routing complexity.
We argue that channel topology evolution should be an autonomous algorithmic
process in order to avoid centralization and to take economic incentives
into account from the very beginning of network bootstrapping.

\begin{acks}
This work was partially funded by the Baden-W\"urttemberg Stiftung and the Federal Ministry of Economic Affairs and Energy on the basis of a decision by the German Bundestag.
%
%
\end{acks}

\bibliographystyle{ACM-Reference-Format}
\bibliography{payment_channels}
\end{document}